\documentclass[a4paper]{article}

\usepackage{INTERSPEECH2022}
\usepackage{xcolor}
\usepackage{multirow}
\usepackage{amsmath}
\usepackage{graphicx}
\usepackage{setspace}
\usepackage{mdwlist}
\usepackage{enumitem}
\usepackage{booktabs}

\DeclareMathOperator{\Tr}{Tr}
\DeclareMathOperator{\softmax}{softmax}

\title{Attention-based conditioning methods using variable frame rate for style-robust speaker verification}
\name{Amber Afshan\thanks{This work was supported in part by the NSF.}, Abeer Alwan}
\address{
  Department of Electrical and Computer Engineering, University of California Los Angeles, USA}
\email{amberafshan@g.ucla.edu, alwan@g.ucla.edu}

\begin{document}

\maketitle
\begin{abstract}
  We propose an approach to extract speaker embeddings that are robust to speaking style variations in text-independent speaker verification. Typically, speaker embedding extraction includes training a DNN for speaker classification and using the bottleneck features as speaker representations. Such a network has a pooling layer to transform frame-level to utterance-level features by calculating statistics over all utterance frames, with equal weighting. However, self-attentive embeddings perform weighted pooling such that the weights correspond to the importance of the frames in a speaker classification task. Entropy can capture acoustic variability due to speaking style variations. Hence, an entropy-based variable frame rate vector is proposed as an external conditioning vector for the self-attention layer to provide the network with information that can address style effects. This work explores five different approaches to conditioning. The best conditioning approach, concatenation with gating, provided statistically significant improvements over the x-vector baseline in 12/23 tasks and was the same as the baseline in 11/23 tasks when using the UCLA speaker variability database. It also significantly outperformed self-attention without conditioning in 9/23 tasks and was worse in 1/23. The method also showed significant improvements in multi-speaker scenarios of SITW. 

\end{abstract}


\section{Introduction}
\label{sec:intro}
Speaking style varies in day-to-day situations such as when reading aloud, talking with a friend, or being highly emotional. These variations result in differences in acoustic properties such as pauses between words, different speaking rates, prosodic variations, elongation of vowels, and incomplete plosive bursts~\cite{picheny_speaking_1986, eskenazi_trends_1993}. Such variability degrades the performance of automatic speaker verification (ASV) systems. The effects of style mismatch on ASV performance were analyzed in~\cite{park_speaker_2016, park_using_2017, park_target_2019, afshan_speaker_2020, afshanspeaker2022}, and some studies addressed such degradation by the use of a joint factor analysis framework~\cite{shriberg_does_2009, chen_compensation_2012}. A recent study~\cite{zhang_analysis_2018} used curriculum-learning-based transfer learning to address style-variability in neutral/physical stress situations. However, the majority of these studies assume the presence of training data in different style conditions. But corpora with different speaking styles from a sufficient number of speakers are not available. 

Our recent work~\cite{afshan_variable_2020} addressed the issue of speaking style variability by using an entropy-based variable frame rate (VFR) technique to perform data augmentation. Entropy inherently captures spectral and temporal variations across different styles~\cite{afshan_variable_2020, afshanspeaking2022, ravi_voice_2019}. Thus, by applying VFR, style-variant speaker embeddings were obtained. However, that work performed data augmentation on the PLDA backend and the speaker embeddings themselves were not style-robust. In contrast, this paper focuses on applying VFR during training of the embedding extractor to obtain embeddings that are style robust.

ASV systems generally use pooling to obtain a fixed-dimension representation from variable-length utterances. In~\cite{variani_deep_2014}, the pooling was performed at the last hidden layer. Recently,~\cite{snyder_deep_2017, snyder_x-vectors_2018} used a statistics pooling layer to calculate the mean and standard deviation of the utterance resulting in a fixed-dimension representation assuming each frame to be equally important. However, we know that not all frames are equally important in conveying speaker or content information~\cite{zhu_use_2000}. To address this issue few studies~\cite{zhang_end--end_2017, zhu_self-attentive_2018, okabe_attentive_2018, desplanques_ecapa-tdnn_2020,  zhu_serialized_2021} have proposed using self-attention in the pooling layer and have observed performance improvements in ASV tasks. Recently,~\cite{wang_attention_2018} decoupled attention weights extracted from an x-vector system and used them in combination with an i-vector system and showed performance improvements. The results confirm that attention weights can better represent the relative importance of each frame irrespective of the underlying embeddings. To learn weights so that the embeddings are style-robust, the attention network needs information to address style effects. 

The proposed method in this paper was inspired by~\cite{margatina_attention-based_2019}, which integrated external linguistic knowledge as a conditioning vector into the self-attention network in three different ways for NLP tasks. Given our prior work on VFR data augmentation~\cite{afshan_variable_2020}, we hypothesize that the VFR output can be used as a conditioning vector for a self-attention network to extract style-robust speaker representations. Hence, this work proposes a new approach that uses entropy-based VFR as a conditioning vector for the self-attentive pooling layer. To the best of our knowledge, this is the first work to perform external knowledge integration in this manner for speaker representation learning.

\section{Proposed Method}
\label{sec:method}
Differences across speaking styles can be categorized into rhythmic variations, including speech rate, long pauses, changes in phoneme duration, boundary articulation, and prosodic variations. However, prosodic variations are associated with speaker identity and disentangling prosody from speech degrades performance. We address the effects of rhythmic variations between styles on ASV performance. 

The proposed method includes self-attentive statistical pooling with VFR conditioning for style-robust speaker verification. This approach uses an x-vector/PLDA framework~\cite{snyder_x-vectors_2018}. The inputs to this system are 30-dimensional mel-frequency cepstral coefficients (MFCCs) using a 25~ms frame length and a 10~ms frame shift. The MFCCs are mean normalized over a sliding window of up to 3~secs. Extrinsic data augmentation of noise and reverberation~\cite{snyder_x-vectors_2018} was applied to the training data.

\subsection{Network architecture}
\label{sec:network_architecture}

\begin{figure}[t]
    \centering
    \includegraphics[scale=0.23]{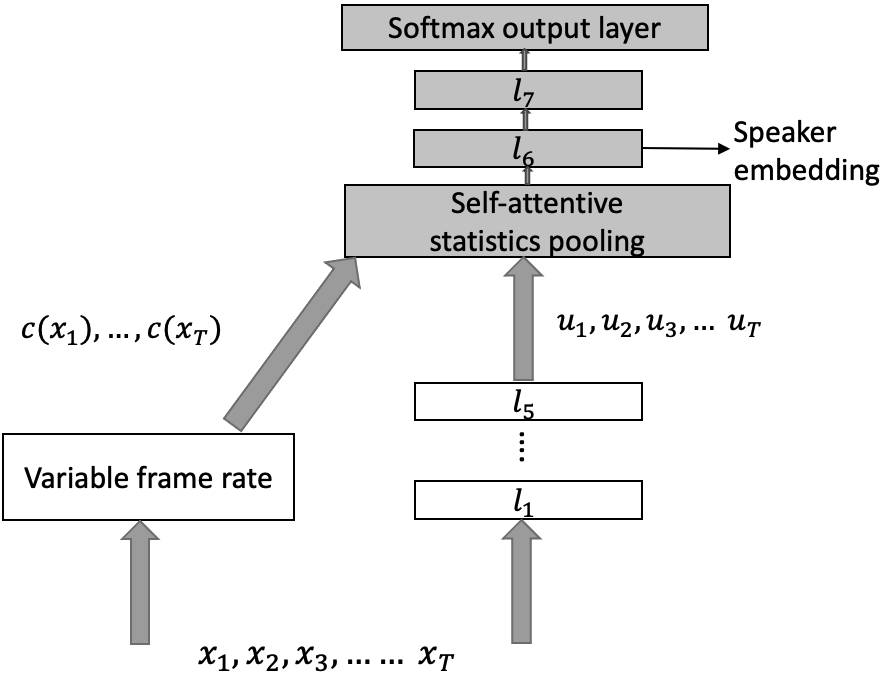}
    \caption{Self-attentive statistics pooling with entropy-based variable frame rate conditioning.}
    \label{fig:block_diagram}
\end{figure}

The network architecture of the proposed method is shown in Figure~\ref{fig:block_diagram}. It builds upon the network structure from x-vectors. Layers $l_1$ to $l_5$ operate at the frame-level, with a small temporal context centered at the current frame $t$. $l_1$ operates on frames $(t-2)$ to $(t+2)$, followed by $l_2$ which operates on the output $l_1$ at time steps \{$t-2, t, t+2$\} and finally  $l_3$ operates on the output of $l_2$ at time steps \{$t-3, t, t+3$\}. Layers $l_4$ and $l_5$ do not add temporal contexts, resulting in a total temporal context of fifteen frames. The pooling layer uses self-attention with conditioning vector, $c(\mathbf{x}_t)$ providing weighted statistics. The output of pooling is propagated to the fully connected layers $l_6$ and $l_7$ and to the softmax output layer. The network is trained to classify speakers using cross-entropy. ReLUs are used as non-linearities. The output of the affine component of $l_6$ is used as speaker embedding and sent to the PLDA backend. 

\subsection{Self-attentive pooling}
\label{ssec:selfattention}
As described earlier, self-attentive pooling learns weights maximizing the speaker classification performance during training and calculates weighted statistics. Let the input to the pooling layer from hidden layer $l_5$ at frame $t$ be $\mathbf{u}_t$. Self-attention~\cite{cheng_long_2016} calculates scores $\alpha_t$ for each frame providing the weighted average ($\Tilde{\mu}$) and the weighted standard deviation ($\Tilde{\sigma}$) of $\mathbf{u}_t$:
\begin{align}
 \label{eq:attention}
     \alpha_t &= \softmax(\mathbf{W}_2^T f(\mathbf{W}_1 \mathbf{u}_t + \mathbf{b}_1) + b_2) \\
    \mathbf{\Tilde{\mu}} = &\sum_{t=1}^T \alpha_t \mathbf{u}_t , \hspace{0.5em}
   \mathbf{ \Tilde{\sigma}} = \sqrt{\sum_{t=1}^T \alpha_t \mathbf{u}_t \odot \mathbf{u}_t -  \mathbf{\Tilde{\mu}} \odot \mathbf{\Tilde{\mu}}}
    \label{eq:sigma}
\end{align}
\noindent  where $\odot$ is the Hadamard product; $\mathbf{W}_2$ and $\mathbf{W}_1$ are the weight matrices and $\mathbf{b}_1$ and $b_2$ are biases  for the attention layer; $f(.)$ is a non-linear activation function, a sigmoid in this case. 

\subsection{External conditioning: Variable frame rate}
\label{ssec:vfr}
In~\cite{afshan_variable_2020}, an entropy-based VFR technique was proposed to generate style-variant representations for data augmentation to address speaking style mismatch. Here, entropy-based VFR output is used as a conditioning vector in the self-attention network.

\subsubsection{Entropy Computation}
\label{sssec:entropy_computation}
Assume a random variable $\mathbf{\nu} \in \mathcal{R}^K$ whose probability distribution function (PDF), $p(\mathbf{\nu})$ is a $K$-dimensional Gaussian. Let the mean and covariance matrix of the random variable be represented as $\mathbf{\mu}$ and $\mathbf{\Sigma}$. The entropy can be calculated as:
   \begin{align}
    H(\mathbf{\nu}) &= - \int p(\mathbf{\nu}) \ln{p(\mathbf{\nu})} d{\mathbf{\nu}} \notag\\
    &= - \int p(\mathbf{\nu}) \Bigg[ -\frac{1}{2}(\mathbf{\nu}-\mathbf{\mu})^T \Sigma^{-1} (\mathbf{\nu}-\mathbf{\mu}) - \ln{|2\pi\mathbf{\Sigma}|^{ \frac{1}{2}}}\Bigg] d{\mathbf{\nu}}\notag\\
     &= \frac{K}{2} + \frac{1}{2} \ln{|2\pi \mathbf{\Sigma}|} \label{eq:eq1}
\end{align} 
 The entropy calculation in Equation~\ref{eq:eq1} is approximated as  $H(\mathbf{\nu}) \approx K \ln{\sqrt{2\pi}} +  \ln{\Tr{\mathbf{\Sigma}}}$, to facilitate faster computation and avoid an ill-posed problem when the covariance matrix of $\nu$ is not full rank~\cite{you_entropy-based_2004}.
   

\subsubsection{\label{vfrcomp}Implementation of variable frame rate}

We use a 25~ms Hamming window and ``oversample'' it with a frame shift of 2.5~ms. The mel-filter spectra are used to calculate the entropy curve every 15~ms with a 30~ms buffer. The frame picking rate is decided based on the entropy curve $H(\mathbf{\nu}_i)$, $i = 1,..., N$. The thresholds ($T_1,T_2, T_3$), calculated as shown in Equation~\ref{eq:thresholds}, are applied to select the frame picking rates:
\begin{equation}
\label{eq:thresholds}
    \begin{cases}
    T_1 &= \omega_1 M_{max} + (1 - \omega_1) M_{med}\\
    T_2 &= (1 - \omega_2)M_{max} + \omega_2M_{med}\\
    T_3 &= (1 - \omega_3)M_{med} + \omega_3M_{min},\\
    \end{cases}
\end{equation}
\noindent where $\omega_1$, $\omega_2,$ and $\omega_3$ are weighting parameters of values 0.7, 0.8, and 0.5, respectively~\cite{you_entropy-based_2004}. $M_{max}$, $M_{med},$ and $M_{min},$ are the maximum, median, and minimum of the entropy curve per utterance, respectively. We create a vector, $z(\mathbf{x})$ composed of 1's and 0's where 1 indicates that the frame is picked and 0 indicates that the frame is skipped. We compare the entropy, $H(\mathbf{\nu}_i)$ with thresholds from Equation~\ref{eq:thresholds} and pick every $r^{th}$ frame from $z(\mathbf{x})$  where $r$ is a multiple of the 2.5~ms frame shift: %
\begin{equation}
\label{eq:rate}
    r = 
\begin{cases}
    2 ,& \text{if }  H(\mathbf{\nu}_i) \geq T_1\\
    3, & \text{if }  T_1 > H(\mathbf{\nu}_i) \geq T_2 \\
    4, & \text{if }  T_2 > H(\mathbf{\nu}_i) \geq T_3 \\
    5, & \text{if } T_3 > H(\mathbf{\nu}_i)
\end{cases}
\end{equation}
When the entropy is high, more frames are selected, and when the entropy is low, fewer frames are selected. Thus, equalizing the entropy across the utterance. This ``oversampled'' (4 times that of MFCCs) vector $z(\mathbf{x})$  is reduced by calculating a sum over every 4 frames to obtain the conditioning vector $c(\mathbf{x})$,
\begin{equation}
        c(\mathbf{x}_{4i}) = \sum_{j=1}^{4} z(\mathbf{x}_{4i+j}) \;,\;\; i={0,1,\dots,\frac{N}{4}} 
\end{equation}
Recall, that we focus on compensating for the rhythmic variations between styles, and these variations can be captured by between-frame entropy. Hence, we hypothesize that an entropy-based conditioning vector may implicitly represent spectral and temporal variations in style and thereby provide self-attention with information to compensate for style effects.

\subsection{Conditional Attention}
\label{ssec:conditionalattention}

As mentioned earlier, VFR is used as a conditioning vector, $c(\mathbf{x}_t)$ in self-attentive pooling by updating $f(\mathbf{u}_t)$ in Equation~\ref{eq:attention} with $f(\mathbf{u}_t,c(\mathbf{x}_t))$. There are multiple ways for adding the conditioning vector and three such methods are explored~\cite{margatina_attention-based_2019}. 

\subsubsection{Conditioning by concatenation}
\label{sssec:concatentation}
The conditioning vector is concatenated with the output of $l_5$, adding extra dimensions to $\mathbf{u}_t$. These new dimensions carry information about the signal's entropy.  $||$ indicates concatenation, $\mathbf{W}_c$ is the weight matrix, and $\mathbf{b}_c$ is the bias vector. Thus:
\begin{equation}
    f_c(\mathbf{u}_t,c(\mathbf{x}_t)) = tanh(\mathbf{W}_c[\mathbf{u}_t||c(\mathbf{x}_t)] + \mathbf{b}_c)
\end{equation}

\subsubsection{Conditioning by gating}
\label{sssec:gating}
A gating mechanism is used to learn a feature mask from $c(\mathbf{x}_t)$  and apply it to $\mathbf{u}_t$ ( $t=1,\dots,T$) the output of the hidden layer ($l_5$) before pooling. A sigmoid is used for the mask to generate values between 0 and 1. As the VFR output conditions gating, frames are selected for pooling based on signal entropy. $\mathbf{W}_g$ is the weight matrix and $\mathbf{b}_g$ is the bias vector. Hence:
\begin{equation}
    f_g(\mathbf{u}_t,c(\mathbf{x}_t)) = \sigma(\mathbf{W}_g(c(\mathbf{x}_t) + \mathbf{b}_g) \odot \mathbf{u}_t
\end{equation}

\subsubsection{Conditioning using affine transformation}
\label{sssec:affine}
An affine transformation is applied on the hidden layer ($l_5$) output, $\mathbf{u}_t$ by using the conditional vector to calculate scaling $\gamma(.)$ and shifting $\beta(.)$. $\mathbf{W}_\gamma, \mathbf{W}_\beta $ are the weight matrices and  $\mathbf{b}_\gamma, \mathbf{b}_\beta$ are the bias vectors, providing self-attention:
\begin{align}
    \label{eq:affine}
    f_a(\mathbf{u}_t,c(\mathbf{x}_t)) &= \gamma(c(\mathbf{x}_t)) \odot \mathbf{u}_t + \beta (c(\mathbf{x}_t))\\
    \gamma(\mathbf{x}) = \mathbf{W}_\gamma & \mathbf{x} + \mathbf{b}_\gamma, \beta(\mathbf{x}) = \mathbf{W}_\beta \mathbf{x} + \mathbf{b}_\beta 
\end{align}
Additional methods are studied: concatenation in combination with gating, and concatenation in combination with affine. Gating is a special case of affine transformation with $\beta=0$ and $\gamma \in [0,1]$. Thus, those two methods are not combined. 

\section{Experiments and Results}
\label{sec:experiments_and_results}

\begin{table*}[t]
\caption{EER(\%) for the UCLA database. Best results for each task are boldfaced. If denoted by a `*' it is not a statistically significant improvement over the baseline. Combined A (concatenation with gating) and Combined B (concatenation with affine).} %
\label{tab:results_ucla}
\centering
\resizebox{0.95\linewidth}{!}{%
\begin{tabular}{c|c|rrr|rrr|rr}
\toprule
\toprule
\textbf{Enroll} &
  \textbf{Test} &
  \textbf{x-vector} &
  \textbf{VFR weights} &
  \textbf{Self-attention} &
  \textbf{Concatenation} &
  \textbf{Gating} &
  \textbf{Affine} &
  \textbf{Combined A} &
  \textbf{Combined B} \\
 \midrule
 \midrule
\multirow{5}{*} {read}   & read         & 0.50 & 1.00    & 0.50 & 0.50 & 0.50 & 0.50 &  0.50 & 0.50 \\
             & instructions & 0.49 & 2.44 & 0.49& 0.49 &  0.49&  0.49 &  0.49 &  0.49\\
             & conversation       & 2.86 & 6.86 & \textbf{2.29} &\textbf{ 2.29 }& 2.86 & \textbf{2.29} &\textbf{ 2.29} & 2.86\\
             & narrative    & 0.80 & 2.55 & 0.80 & 0.80 & 0.80 & \textbf{0.64* }&  0.80 & \textbf{0.64*}\\
             & pet-directed          & 17.14& 22.86& 17.14& \textbf{14.29}& 17.14& \textbf{14.29} &  \textbf{14.29} &  \textbf{14.29}\\
             \midrule
\multirow{5}{*} {instructions} & read         & 1.47 & 3.43 & 1.47 & \textbf{0.98} & 1.47 &  1.47 & \textbf{0.98} &  1.47 \\
             & instructions & 0.45 & 0.45 & 0.45 & 0.45 & 0.45 & 0.45 &  0.45 & 0.45 \\
             & conversation       & 2.79 & 6.70 & 2.79 & 2.79 & 3.35 &  3.35 &  2.79  &  \textbf{2.24}\\
             & narrative    & 1.23 & 2.61 & 1.08 & 0.92 & 0.92 & 0.92 & \textbf{ 0.77} & 0.92\\
             & pet-directed          & 18.92& 24.32& 16.22& 16.22& 16.22& 16.22& \textbf{13.51} & 16.22\\
             \midrule
\multirow{5}{*} {conversation} & read         & 2.03 & 5.08 & \textbf{1.52} & \textbf{1.52 }& 2.03 & \textbf{1.52} & \textbf{1.52} & \textbf{1.52} \\
             & instructions & 2.97 & 4.95 & \textbf{2.48*} & \textbf{2.48*} & 2.97 & \textbf{2.48*} & \textbf{ 2.48*} & \textbf{ 2.48*}\\
             & conversation       & 0.57 & 1.72 & 0.57 & 0.57 & 0.57 & 0.57 & 0.57 & 0.57\\
             & narrative    & 1.94 & 5.98 & \textbf{1.78} & 1.94 & 2.59 &  2.10 & 1.94&  2.10  \\
             & pet-directed          & 20.00   & 22.86& 20.00   & 20.00   & \textbf{17.14}& \textbf{17.14}&  \textbf{17.14}& \textbf{17.14}\\
             \midrule

\multirow{5}{*} {narrative}    & read         & 0.48 & 1.76 &0.32* &0.32* & 0.32* & 0.32* &  0.32* & \textbf{0.16}\\
             & instructions & 0.46 & 1.08 & 0.46 & 0.46 & 0.46 &0.46&  0.46 &  0.46\\
             & conversation       & 1.46 & 4.56 & 1.64 & 1.83 & 1.64 & 1.28 & \textbf{1.10}  & \textbf{1.10}\\
             & pet-directed          & 18.58& 26.55& 18.58&\textbf{ 13.27}& 18.58& 15.93 & \textbf{13.27} & 16.81\\
             \midrule

\multirow{5}{*} {pet-directed} & read         & 14.29& 20.00   & 14.29& \textbf{11.43}& 14.29& 14.29&  14.29 &  14.29\\
             & instructions & 18.92& 27.03& 18.92& 18.92& 16.22& 16.22 & \textbf{13.51} & 16.22\\
             & conversation       & 21.21& 24.24& 21.21& 21.21& \textbf{18.18}&  21.21 & \textbf{18.18} & 21.21\\
             & narrative    & 19.47& 28.32& 17.70& 15.93& 19.47& 17.70 & \textbf{14.16} & 17.70\\
             \bottomrule
             \bottomrule
\end{tabular}%
}
\end{table*}

\begin{table}[t]
\caption{EER(\%) for the SITW database. Best results for each task are boldfaced and are statistically significant improvement over the baseline. Combined A (concatenation with gating) and Combined B (concatenation with affine).
}
\label{tab:sitw_eval}
\centering
\resizebox{\linewidth}{!}{%
\begin{tabular}{c|rrrr}
\toprule
\toprule
\textbf{Model }            & \textbf{Core-Core} & \textbf{Core-Multi }& \textbf{Assist-Core} & \textbf{Assist-Multi} \\
\midrule
\midrule
x-vector   &         3.66     & 5.87       & 5.47        & 6.9          \\
VFR weights       & 8.17      & 10.67      & 10.17       & 11.82        \\
Self-attention    & 3.91      & 6.09       & 5.51        & 6.64         \\
\hline
Concatenation      & 3.86      & 5.83       & 5.35        &\textbf{6.35 }        \\
Gating            & 4.32      & 6.64       & 6.25        & 7.61         \\
Affine            & 3.91      & 5.95       & 5.41        & 6.58         \\
\hline
Combined A        & 3.69      &\textbf{5.81}       &\textbf{5.26 }       & 6.54        \\
Combined B        & 3.91      & 6.17       & 5.83        & 7.16 \\
\bottomrule
\bottomrule

\end{tabular}%
}
\end{table}

\subsection{Databases}
\label{ssec:data}

\subsubsection{The UCLA Speaker Variability Database (SVD)}

The UCLA SVD database~\cite{keating_new_2019, kreiman_relationship_2015, keating_ldc_2021,keating_ucla_2021} is used to evaluate ASV performance in the presence of style variability. It consists of recordings from 202 female and male speakers performing various speech tasks in a sound-attenuated booth with a sampling rate of 22kHz. We used \textbf{reading} sentences ($\approx 75$ sec per speaker); giving \textbf{instructions} representing unscripted clear monologue style ($\approx 30$ sec  per speaker); \textbf{narrating} a recent neutral, happy, or annoying conversation representing unscripted affective speech ($\approx$ 30 sec each affect per speaker); speaker's side of the conversation on a call representing unscripted \textbf{conversational} style (60--120 sec  per speaker); and \textbf{pet-directed} speech, characterized by exaggerated prosody (60--120 sec  per speaker). 
To effectively evaluate style-robustness, we require negligible effect from phonetic variability. We use 30-sec long speech samples to cover enough phonetic variability and capture speaker idiosyncratic information~\cite{hasan_duration_2013}. A total of 1,838 30-sec segments are extracted. We require a minimum of 1~min of speech per speaker to generate style-matched trials. However, many speakers had less than 1~min of pet-directed or affect-matched narrative speech, and style-matched tasks for those styles were omitted. We hence obtain 23 tasks (5 styles in matched and mismatched conditions except for style-matched pet-directed and narrative cases). To match the sampling rate of other databases used, the data were downsampled to 16~kHz.

\subsubsection{The Speakers in the Wild Database (SITW)}
The proposed approach was also evaluated on SITW~\cite{mclaren_speakers_2016} \textit{EVAL} set to examine its effects on a large-scale public database.
SITW has 2,883 recordings from 117 male and 63 female speakers divided into 6,445 utterances sampled at 16~kHz. SITW consists of both single- and multiple-speaker audio with segment labels for person of interest (POI) in enrollment utterances. Enrollment utterances include core (single POI) and assist (multiple speakers with segmentation labels for POI) and test utterances include core (single POI) and multi (multiple speakers with no segmentation labels for POI). 
\subsubsection{VoxCeleb Database} 
Training was performed using the \textit{DEV} set from Voxceleb2~\cite{chung_voxceleb2_2018}. It consists of speech from YouTube videos of 3,682 male and 2,313 female speakers and includes 1,092,009 utterances with a sampling rate of 16~kHz. We did not use VoxCeleb2 for testing because it comprises interview-style speech and does not include different styles per speaker. 

\subsection{Experimental Setup}
\label{ssec:experimental_setup}
As is commonly used, the embedding extractor had 512 nodes in each of $l_1$ to $l_4$, $l_5$ had 1500 nodes, while $l_6$ and $l_7$ had 512 nodes. The self-attention layer had 500 nodes. The experiments were setup using Pytorch~\cite{paszke_pytorch_2019} and Kaldi~\cite{povey_kaldi_2011} with Adam optimizer~\cite{kingma_adam_2017} and a batch size of 128 trained for 100 epochs. 

\subsection{Results and Discussion}
\label{ssec:results_and_discussion}

\subsubsection{UCLA SVD evaluation}
\label{sssec:ucla_eval}
Table~\ref{tab:results_ucla} shows the equal error rate (EER) for an x-vector baseline, self-attentive network, and the five conditioning methods.  Statistical significance ($p<0.05$) was evaluated using the McNemar's test~\cite{mcnemar_note_1947}. The baseline x-vector performs better for style-matched tasks than style-mismatched ones. For instance, when enrolled with conversational speech, the style-matched task results in an EER of 0.57\%. However, style-mismatched tasks have EERs of 2.03\%, 2.97\%, 1.94\% and 20\% for read, instructions, narrative, and pet-directed speech, respectively.


To evaluate the need for a self-attention layer, another model is trained with VFR as weights for statistical pooling. However, as seen in Table~\ref{tab:results_ucla}, performance degrades when using this approach (VFR weights). Thus, VFR by itself may not be sufficient to provide meaningful weights for each frame. 

Self-attentive speaker embeddings provide a statistically significant improvement over x-vector baseline in 6/23 tasks and only degrades in the narrative--conversation task. These improvements are due to embeddings with better speaker discrimination capabilities, in agreement with the results in ~\cite{zhu_self-attentive_2018, okabe_attentive_2018}.

Compared to the x-vector performance, among the proposed approaches of VFR conditioning, Combined A (concatenation with gating) results in statistically significant improvements in 12/23 tasks, while Combined B (concatenation with affine transformation) results in statistically significant improvements in 6/23 tasks. Among the three individual VFR conditioning methods, concatenation results in statistically significant improvements in 8/23 tasks, gating in 5/23 tasks, and affine transformation in 10/23 tasks. For the remaining tasks, all conditioning methods perform the same as x-vector. Gating is a special case of affine transformation, and individually gating performs worse than affine, but when combined with concatenation, it performs better. Moreover, the best performing method, VFR conditioning by concatenation with gating, provided significant improvement over the self-attentive embeddings in 10/23 tasks and only degraded in conversation--narrative tasks.  The results support the hypothesis that including the VFR conditioning vector in self-attention facilitates the speaker representations to be robust to speaking style variations. 

In all style-matched cases i.e, for read--read, instructions--instructions and conversation--conversation, no performance improvement was observed. Because there are no style differences in these cases, it is expected that there may not be improvements for style-matched cases.  Among the style-mismatched cases, performance remained the same for read--instruction and conversation--narrative. Read and instruction are closely-related styles as one is scripted clear speech and the other is an unscripted clear monologue speech. In fact, the performance in the case of read--instruction is close to the style-matched case of read--read. We believe the same applies to conversation and narrative, because these two styles are closely related, one being unscripted conversational speech and the other being unscripted narration. All other cases improved in at least one of the conditioning approaches. Attention visualization is not presented as it did not provide an intuitive explanation.

\subsubsection{SITW evaluation}
\label{sssec:sitw_eval}
SITW evaluation results are in Table~\ref{tab:sitw_eval}. Conditioning provides improvements over x-vectors in Core-Multi, Assist-Core, and Assist-Multi. For the Core-Multi and Assist-Core cases, the best performing method is conditioning using concatenation with gating. However, in Assist-Multi, conditioning with concatenation performs the best. The proposed method provides performance improvements in multi-speaker scenarios. These scenarios include more variations in style as they are dialogues. 

\section{Conclusion}
\label{sec:conclusion}
This paper shows that entropy-based VFR used to condition self-attentive speaker embeddings provide style-robust representations. The best conditioning approach, concatenation with gating, results in statistically significant ASV improvements over the x-vector baseline for both the UCLA SVD database and multi-speaker scenarios in the SITW evaluation set. In the future, we will investigate the proposed method on short-duration scenario~\cite{guo_speaker_2016, ravi_exploring_2020} and with other embedding extractors that utilize a pooling layer~\cite{zhou_resnext_2021, desplanques_ecapa-tdnn_2020}.




\bibliographystyle{IEEEtran}

\bibliography{refs}

\begin{thebibliography}{10}
\providecommand{\url}[1]{#1}
\csname url@samestyle\endcsname
\providecommand{\newblock}{\relax}
\providecommand{\bibinfo}[2]{#2}
\providecommand{\BIBentrySTDinterwordspacing}{\spaceskip=0pt\relax}
\providecommand{\BIBentryALTinterwordstretchfactor}{4}
\providecommand{\BIBentryALTinterwordspacing}{\spaceskip=\fontdimen2\font plus
\BIBentryALTinterwordstretchfactor\fontdimen3\font minus
  \fontdimen4\font\relax}
\providecommand{\BIBforeignlanguage}[2]{{%
\expandafter\ifx\csname l@#1\endcsname\relax
\typeout{** WARNING: IEEEtran.bst: No hyphenation pattern has been}%
\typeout{** loaded for the language `#1'. Using the pattern for}%
\typeout{** the default language instead.}%
\else
\language=\csname l@#1\endcsname
\fi
#2}}
\providecommand{\BIBdecl}{\relax}
\BIBdecl

\bibitem{picheny_speaking_1986}
M.~A. Picheny, N.~I. Durlach, and L.~D. Braida, ``Speaking clearly for the hard
  of hearing {II}: {Acoustic} characteristics of clear and conversational
  speech,'' \emph{JSLHR}, vol.~29, no.~4, pp. 434--446, 1986.

\bibitem{eskenazi_trends_1993}
M.~Eskenazi, ``Trends in speaking styles research,'' in \emph{EUROSPEECH},
  1993.

\bibitem{park_speaker_2016}
S.~J. Park, C.~Sigouin, J.~Kreiman, P.~A. Keating, J.~Guo, G.~Yeung, F.-Y. Kuo,
  and A.~Alwan, ``Speaker {Identity} and {Voice} {Quality}: {Modeling} {Human}
  {Responses} and {Automatic} {Speaker} {Recognition},'' in \emph{Interspeech},
  2016.

\bibitem{park_using_2017}
S.~J. Park, G.~Yeung, J.~Kreiman, P.~A. Keating, and A.~Alwan, ``Using {Voice}
  {Quality} {Features} to {Improve} {Short}-{Utterance}, {Text}-{Independent}
  {Speaker} {Verification} {Systems},'' \emph{Interspeech}, 2017.

\bibitem{park_target_2019}
S.~J. Park, A.~Afshan, J.~Kreiman, G.~Yeung, and A.~Alwan,
  ``\BIBforeignlanguage{en}{Target and {Non}-target {Speaker} {Discrimination}
  by {Humans} and {Machines}},'' in \emph{\BIBforeignlanguage{en}{ICASSP}},
  2019, pp. 6326--6330.

\bibitem{afshan_speaker_2020}
A.~Afshan, J.~Kreiman, and A.~Alwan, ``Speaker discrimination in humans and
  machines: {Effects} of speaking style variability,'' in \emph{{Interspeech}},
  2020.

\bibitem{afshanspeaker2022}
------, ``Speaker discrimination performance for ``easy'' versus ``hard''
  voices in style-matched and -mismatched speech,'' \emph{J. Acoust. Soc. Am.},
  vol. 151, no.~2, pp. 1393--1403, 2022.

\bibitem{shriberg_does_2009}
E.~Shriberg, S.~Kajarekar, and N.~Scheffer, ``Does session variability
  compensation in speaker recognition model intrinsic variation under
  mismatched conditions?'' in \emph{{Interspeech}}, 2009.

\bibitem{chen_compensation_2012}
S.~Chen and M.~Xu, ``Compensation of {Intrinsic} {Variability} with {Factor}
  {Analysis} {Modeling} for {Robust} {Speaker} {Verification},'' in
  \emph{{Interspeech}}, 2012.

\bibitem{zhang_analysis_2018}
C.~Zhang, S.~Ranjan, and J.~H. Hansen, ``An {Analysis} of {Transfer} {Learning}
  for {Domain} {Mismatched} {Text}-independent {Speaker} {Verification}.'' in
  \emph{Odyssey}, 2018, pp. 181--186.

\bibitem{afshan_variable_2020}
A.~Afshan, J.~Guo, S.~J. Park, V.~Ravi, A.~McCree, and A.~Alwan, ``Variable
  frame rate-based data augmentation to handle speaking-style variability for
  automatic speaker verification,'' in \emph{Interspeech}, Shanghai, China,
  2020, pp. 4318--4322.

\bibitem{afshanspeaking2022}
A.~Afshan, ``\BIBforeignlanguage{en}{Speaking style variability in speaker
  discrimination by humans and machines},'' Ph.{D}. Dissertation, University of
  California, Los Angeles, CA, 2022.

\bibitem{ravi_voice_2019}
V.~Ravi, S.~J. Park, A.~Afshan, and A.~Alwan, ``Voice {Quality} and
  {Between}-{Frame} {Entropy} for {Sleepiness} {Estimation}.'' in
  \emph{Interspeech}, Graz, Austria, 2019, pp. 2408--2412.

\bibitem{variani_deep_2014}
E.~Variani, X.~Lei, E.~McDermott, I.~L. Moreno, and J.~Gonzalez-Dominguez,
  ``Deep neural networks for small footprint text-dependent speaker
  verification,'' in \emph{{ICASSP}}, 2014, pp. 4052--4056, iSSN: 2379-190X.

\bibitem{snyder_deep_2017}
D.~Snyder, D.~Garcia-Romero, D.~Povey, and S.~Khudanpur, ``Deep {Neural}
  {Network} {Embeddings} for {Text}-{Independent} {Speaker} {Verification}.''
  in \emph{Interspeech}, 2017, pp. 999--1003.

\bibitem{snyder_x-vectors_2018}
D.~Snyder, D.~Garcia-Romero, G.~Sell, D.~Povey, and S.~Khudanpur, ``X-vectors:
  {Robust} dnn embeddings for speaker recognition,'' in \emph{{ICASSP}}, 2018.

\bibitem{zhu_use_2000}
Q.~Zhu and A.~Alwan, ``On the use of variable frame rate analysis in speech
  recognition,'' in \emph{{ICASSP}}, vol.~3.\hskip 1em plus 0.5em minus
  0.4em\relax IEEE, 2000, pp. 1783--1786.

\bibitem{zhang_end--end_2017}
S.-X. Zhang, Z.~Chen, Y.~Zhao, J.~Li, and Y.~Gong, ``End-to-{End} {Attention}
  based {Text}-{Dependent} {Speaker} {Verification},'' in
  \emph{{arXiv}:1701.00562 [cs, stat]}.

\bibitem{zhu_self-attentive_2018}
Y.~Zhu, T.~Ko, D.~Snyder, B.~Mak, and D.~Povey,
  ``\BIBforeignlanguage{en}{Self-{Attentive} {Speaker} {Embeddings} for
  {Text}-{Independent} {Speaker} {Verification}},'' in
  \emph{\BIBforeignlanguage{en}{Interspeech}}, 2018, pp. 3573--3577.

\bibitem{okabe_attentive_2018}
K.~Okabe, T.~Koshinaka, and K.~Shinoda, ``Attentive {Statistics} {Pooling} for
  {Deep} {Speaker} {Embedding},'' \emph{Interspeech}, pp. 2252--2256, 2018.

\bibitem{desplanques_ecapa-tdnn_2020}
B.~Desplanques, J.~Thienpondt, and K.~Demuynck, ``{ECAPA}-{TDNN}: {Emphasized}
  {Channel} {Attention}, {Propagation} and {Aggregation} in {TDNN} {Based}
  {Speaker} {Verification},'' in \emph{Interspeech}, 2020, pp. 3830--3834.

\bibitem{zhu_serialized_2021}
``Serialized {Multi}-{Layer} {Multi}-{Head} {Attention} for {Neural} {Speaker}
  {Embedding},'' 2021, arXiv:2107.06493 [cs, eess].

\bibitem{wang_attention_2018}
Q.~Wang, K.~Okabe, K.~A. Lee, H.~Yamamoto, and T.~Koshinaka, ``Attention
  {Mechanism} in {Speaker} {Recognition}: {What} {Does} it {Learn} in {Deep}
  {Speaker} {Embedding}?'' in \emph{SLT}, 2018, pp. 1052--1059.

\bibitem{margatina_attention-based_2019}
K.~Margatina, C.~Baziotis, and A.~Potamianos, ``Attention-based {Conditioning}
  {Methods} for {External} {Knowledge} {Integration},'' \emph{arXiv:1906.03674
  [cs, stat]}, 2019.

\bibitem{cheng_long_2016}
J.~Cheng, L.~Dong, and M.~Lapata, ``Long {Short}-{Term} {Memory}-{Networks} for
  {Machine} {Reading},'' in \emph{EMNLP}, 2016, pp. 551--561.

\bibitem{you_entropy-based_2004}
H.~You, Q.~Zhu, and A.~Alwan, ``Entropy-based variable frame rate analysis of
  speech signals and its application to {ASR},'' in \emph{{ICASSP}},
  vol.~1.\hskip 1em plus 0.5em minus 0.4em\relax IEEE, 2004, pp. I--549.

\bibitem{keating_new_2019}
P.~Keating, J.~Kreiman, and A.~Alwan, ``\BIBforeignlanguage{en}{A {New}
  {Speech} {Database} {For} {Within}- and {Between}-{Speaker} {Variability}},''
  \emph{\BIBforeignlanguage{en}{Proc of the 19th ICPhS}}, p.~4, 2019.

\bibitem{kreiman_relationship_2015}
J.~Kreiman, S.~J. Park, P.~A. Keating, and A.~Alwan, ``The {Relationship}
  {Between} {Acoustic} and {Perceived} {Intraspeaker} {Variability} in {Voice}
  {Quality},'' in \emph{Interspeech}, Dresden, Germany, 2015.

\bibitem{keating_ldc_2021}
\BIBentryALTinterwordspacing
P.~Keating, J.~Kreiman, A.~Alwan, A.~Chong, and Y.~Lee, ``{UCLA} speaker
  variability database,'' 2021. [Online]. Available:
  \url{https://catalog.ldc.upenn.edu/LDC2021S09}
\BIBentrySTDinterwordspacing

\bibitem{keating_ucla_2021}
\BIBentryALTinterwordspacing
------, ``{UCLA} {Speaker} {Variability} {Database},'' 2021, last viewed
  October 5, 2021. [Online]. Available:
  \url{http://www.seas.ucla.edu/spapl/shareware.html\#Data}
\BIBentrySTDinterwordspacing

\bibitem{hasan_duration_2013}
T.~Hasan, R.~Saeidi, J.~H. Hansen, and D.~A. Van~Leeuwen, ``Duration mismatch
  compensation for i-vector based speaker recognition systems,'' in
  \emph{{ICASSP}}.\hskip 1em plus 0.5em minus 0.4em\relax IEEE, 2013, pp.
  7663--7667.

\bibitem{mclaren_speakers_2016}
M.~McLaren, L.~Ferrer, D.~Castan, and A.~Lawson, ``The {Speakers} in the {Wild}
  ({SITW}) speaker recognition database.'' in \emph{Interspeech}, 2016, pp.
  818--822.

\bibitem{chung_voxceleb2_2018}
\BIBentryALTinterwordspacing
J.~S. Chung, A.~Nagrani, and A.~Zisserman, ``{VoxCeleb2}: {Deep} {Speaker}
  {Recognition},'' \emph{Interspeech 2018}, pp. 1086--1090, Sep. 2018.
  [Online]. Available: \url{http://arxiv.org/abs/1806.05622}
\BIBentrySTDinterwordspacing

\bibitem{paszke_pytorch_2019}
\BIBentryALTinterwordspacing
A.~Paszke, S.~Gross, F.~Massa, A.~Lerer, J.~Bradbury, G.~Chanan, T.~Killeen,
  Z.~Lin, N.~Gimelshein, L.~Antiga, A.~Desmaison, A.~Köpf, E.~Yang, Z.~DeVito,
  M.~Raison, A.~Tejani, S.~Chilamkurthy, B.~Steiner, L.~Fang, J.~Bai, and
  S.~Chintala, ``{PyTorch}: {An} {Imperative} {Style}, {High}-{Performance}
  {Deep} {Learning} {Library},'' \emph{arXiv:1912.01703 [cs, stat]}, Dec. 2019,
  arXiv: 1912.01703. [Online]. Available: \url{http://arxiv.org/abs/1912.01703}
\BIBentrySTDinterwordspacing

\bibitem{povey_kaldi_2011}
D.~Povey, A.~Ghoshal, G.~Boulianne, L.~Burget, O.~Glembek, N.~Goel,
  M.~Hannemann, P.~Motlicek, Y.~Qian, P.~Schwarz, and {others}, ``The {Kaldi}
  speech recognition toolkit,'' IEEE SP Society, Tech. Rep., 2011.

\bibitem{kingma_adam_2017}
\BIBentryALTinterwordspacing
D.~P. Kingma and J.~Ba, ``Adam: {A} {Method} for {Stochastic} {Optimization},''
  \emph{arXiv:1412.6980 [cs]}, Jan. 2017, arXiv: 1412.6980. [Online].
  Available: \url{http://arxiv.org/abs/1412.6980}
\BIBentrySTDinterwordspacing

\bibitem{mcnemar_note_1947}
Q.~McNemar, ``Note on the sampling error of the difference between correlated
  proportions or percentages,'' \emph{Psychometrika}, vol.~12, no.~2, pp.
  153--157, 1947, publisher: Springer.

\bibitem{guo_speaker_2016}
J.~Guo, G.~Yeung, D.~Muralidharan, H.~Arsikere, A.~Afshan, and A.~Alwan,
  ``Speaker {Verification} {Using} {Short} {Utterances} with {DNN}-{Based}
  {Estimation} of {Subglottal} {Acoustic} {Features}.'' in
  \emph{{Interspeech}}, 2016, pp. 2219--2222.

\bibitem{ravi_exploring_2020}
V.~Ravi, R.~Fan, A.~Afshan, H.~Lu, and A.~Alwan, ``Exploring the {Use} of an
  {Unsupervised} {Autoregressive} {Model} as a {Shared} {Encoder} for
  {Text}-{Dependent} {Speaker} {Verification},'' in \emph{Interspeech}, 2020.

\bibitem{zhou_resnext_2021}
T.~Zhou, Y.~Zhao, and J.~Wu, ``{ResNeXt} and {Res2Net} {Structures} for
  {Speaker} {Verification},'' in \emph{2021 {IEEE} {Spoken} {Language}
  {Technology} {Workshop} ({SLT})}, Jan. 2021, pp. 301--307.

\end{thebibliography}

\end{document}